# Connection errors in networks of linear features and the application of geometrical reduction in spatial data algorithms.

Panteleimon Rodis

Diploma in Computer Science, Hellenic Open University.

rodispantelis@gmail.com

pantelisrodis.blogspot.com

**Abstract** We present a study on connection errors in networks of linear features and methods of error detection. We model networks with special connection specifications as networks with hierarchically connected features and define errors considering the spatial relationships and the functionality of the network elements. A general definition of the problem of the detection of connection errors which takes into account the functionality of the network elements is discussed. Then a series of spatial algorithms that solve different aspects of the problem is presented. We also define and analyze the notion of geometrical reduction as a method of achieving efficient performance. In the last section the undecidability of algorithmic error correction is discussed.

**Keywords** Connection error detection, Geometrical reduction, Hierarchical network, Spatial algorithms.

## 1 Introduction

Many types of networks have specifications on the way their linear features should be or should not be connected. Network features according to their state or functionality are required or prohibited to form connections with features of certain properties. Examples of networks of this type are hierarchical networks, where connection specifications define a hierarchy of the network elements. Another example may be found in road networks where high capacity roads and expressways should be interconnected as in any other case traffic problems occur. Specifications of these types may also be found in telecommunication networks, when certain communication channels should never be connected with each other for security of functional reasons. In any mapping process of such a network in Geographical Information Systems (GIS) there can be found errors in the cases that the connection specifications or the topology of the networks are violated. This violation can happen because of hand-made mistakes in the mapping process or because in reality some elements might be connected in a not legal way. In networks mapped in GIS these and any other similar case must be able to be detected efficiently.

Modern GIS software has focused on the development of methods and tools that detect topological relationships and connectivity consistency. Examples of such tools in important GIS products are in ArcGIS (ESRI, 2005a and 2010) and in AutoCAD Map (Autodesk, 2010). This approach is very useful but not enough, the errors that can be found in datasets are not only topological but also functional and this is the main subject that this paper focuses on. The hierarchical structure we use to model connection specifications clearly models the functional properties of the network. The error detection methods we can find in relevant literature do not cover the subject of connection errors as we define it, indicatively we may refer to the paper of Peng Gong and Lan Mu (Gong and Mu, 2000) that briefly indicates the fields that research focuses on.

In §2 we discuss the topological issues and the spatial relationships of networks of linear features, their determination is a critical issue of the paper. On this purpose, we define an object model for spatial objects of various dimensions and we use it to model the self-intersection instances of linear features. Then we define the type of network that is studied in this paper. The notions described by the object and intersection models are used in the development of the algorithms we present.

Before the analysis of the problem of error detection it is necessary to present in §3 the notion of geometrical reduction, as it is a key feature for most algorithms in the paper. The processes that are based on geometrical reduction exploit the spatial relationships among the elements of a dataset as they are defined in §2.

In §4 we define a hierarchical structure for the studied networks and model the connection specifications of network elements. Then we define a general case of the problem of detection of connection errors and present reductions of more special cases that describe different aspects of the problem to the general case.

In the next sections a series of algorithms are presented that solve general and special cases of the problem. For the general case of the problem two algorithms are presented, an exhaustive search algorithm and a more efficient algorithm that uses geometrical reduction. The exhaustive search algorithm uses commonly used data structures and we use it as a mean of comparison to the more efficient algorithms of the paper. Then we present Self-Intersection and flow error detection algorithms that also use geometrical reduction to achieve better efficiency. One important difference among the algorithms is that the exhaustive search algorithm uses the attribute table for the computation of the problem, while the rest of the algorithms use the geometry of the features.

In §8 the complexity analysis of the presented algorithms is discussed, this shows the advantages of applying the notion of geometrical reduction in spatial algorithms.

In the last section we prove that the algorithmic correction of map errors or software bugs is not possible.

## 2 Topological issues and spatial relationships

### 2.1 Object model

Before discussing the problem of connection errors we need to define spatially the network elements and their topological relationships and also the errors that can occur due to invalid topology. There are two types of objects that represent entities necessary to define a network these are points and lines.

The model we propose is based on the notion that the boundaries of a spatial object can be modeled as a graph of vertices and edges. Also, the interior of an $n$-dimensional object is the $n$-dimensional space enclosed by the object boundaries. This is considered in compliance to the point-set model (Egenhofer and Franzosa, 1991), since each edge of an object can also be considered as a set of points.

The objects we define have similar structure with object models like the object representations in the well-known text (WKT) and binary formats (WKB), see (Herring, 2010).

Let us then define spatial objects as sequences of vertices and sets of edges.

A vertex $v_i$ in $n$-dimensional space is a vector of $n$ dimensions, $n \in N^*$. An edge (or arc) is the linear feature that connects two vertices.

- Point is a 0-dimensional object consisting of one vertex $\{v_1\}$.
- Line is an 1-dimensional object consisting of a finite sequence of $|m|$ vertices $(v_1,...., v_m)$, $m \geq 2$ connected by a set of $|m-1|$ distinct edges $\{v_i, v_{i+1}\}$. Where $v_1$ and $v_m$ are the boundaries of the object.
- Polygon is a 2-dimensional object with boundaries consisting of a finite sequence of $|u|$ vertices $(v_1,...., v_u)$, $u \geq 3$ connected by a set of $|u|$ distinct edges $\{v_i, v_{i+1}\} \cup \{v_u, v_1\}$, where $1 \leq i < u$.
- Polyhedron is a 3-dimensional object with boundaries consisting of a finite sequence of $|y|$ vertices $(v_1,...., v_y)$, $y \geq 4$, that are connected by a set of $|t|$ distinct edges $\{v_i, v_{i+1}\}$, where $1 \leq i < t$ and $t > 4$. The degree of each vertex $v_i$ is $\deg(v_i) \geq 3$ and the Euler characteristic $x = 2$.

We defined the vertices of an object as a sequence and not as a set. The advantage of this approach is that it defines the direction of each edge and the way that the vertices are connected. It also defines more accurately the geometry of the object and may also describe complex objects and objects with self-intersecting parts. The edges that form the boundary of an object must be distinct, every pair of vertices must be connected by at most one edge otherwise topological problems may occur.

Networks in GIS are usually defined in $\Re^2$ space. Since the algorithms in this paper are tested in $\Re^2$ space, we define the network in $\Re^2$ space and describe the network elements with simple line features and the connection points with simple point features. There is no real restriction in the dimension of the space that the network can be defined, so the same models and algorithms can be used in multidimensional spaces. We define a network as a collection of sets of vertices and edges that form points and arcs.

- A vertex $v_i$ that denotes a point is a vector $<x_i, y_i>$, where $x_i, y_i \in \Re$. It is defined by the equation $y = ax + b$, for $x = x_i$ and $y = y_i$.
- An edge as a component of a linear feature (arc) is defined by the two vertices $v_1 <x_1, y_1>$ and $v_2 <x_2, y_2>$ that it connects. It is also described by the algebraic equation $y = ax + b$, where $x_1 \leq x \leq x_2$ and $y_1 \leq y \leq y_2$.

## 2.2 Intersection models

The 9-Intersection model (Egenhofer et al, 1993), following the 4-Intersection model, is commonly used for describing spatial relationships. The spatial relationships of the networks that this paper discusses may be described using the 9-Intersection model and a description of self-intersection for linear features.

**9-Intersection model**

| Object $A$ | Object $B$ |
|---|---|
| $A$ (interior) | $B$ (interior) |
| $\partial A$ (boundary) | $\partial B$ (boundary) |
| $A^-$ (exterior) | $B^-$ (exterior) |

$$I_9(A,B) = \begin{Bmatrix} A \cap B & A \cap \partial B & A \cap B^- \\ \partial A \cap B & \partial A \cap \partial B & \partial A \cap B^- \\ A^- \cap B & A^- \cap \partial B & A^- \cap B^- \end{Bmatrix}$$

The 9-Intersection model does not include self-intersection that may occur in spatial objects. Self-intersection is many times forbidden in networks and cannot be overlooked, so it must be defined as an extension of the model. A point object cannot self-intersect, so we define the line self-intersection model.

**Line self-intersection model**

A line self-intersects when two edges that do not connect consecutive vertices intersect; the system of their equations has one or more common solutions. When the common solutions are vertices the line self-intersects over vertices. The line must consist of more than 3 vertices, otherwise self-intersection is not possible.

| Line object $A$ |
|---|
| $A$ (interior) |
| $\partial A$ (boundary) |

- SI($\partial A$). The system of equations of the edges $\{v_1, v_2\}$ and $\{v_{m-1}, v_m\}$ has one solution in point $p \subseteq \partial A$, $p \cap v_1 \neq \emptyset$ and $p \cap v_m \neq \emptyset$.
- SI(A). The system of equations of the edges $E_i \subset A$, $E_j \subset A$ has at least one solution in point $p \not\subset \partial A$.

- SI(∂A, A). Given an edge $E_k$ and a set $I$ of two points $I = \{p_1, p_2\}$, where $I \subset E_k$ and $I \subset A$ and $I \not\subset \partial A$. The system of equations of a line $l$ defined by $I$ and a set of points $B \subseteq \partial A$ has at least one solution and at most two.

$$SI_L(A) = \{SI(\partial A) \quad SI(A) \quad SI(\partial A, A)\}$$

Line self-intersection instances are illustrated in Figure 1.

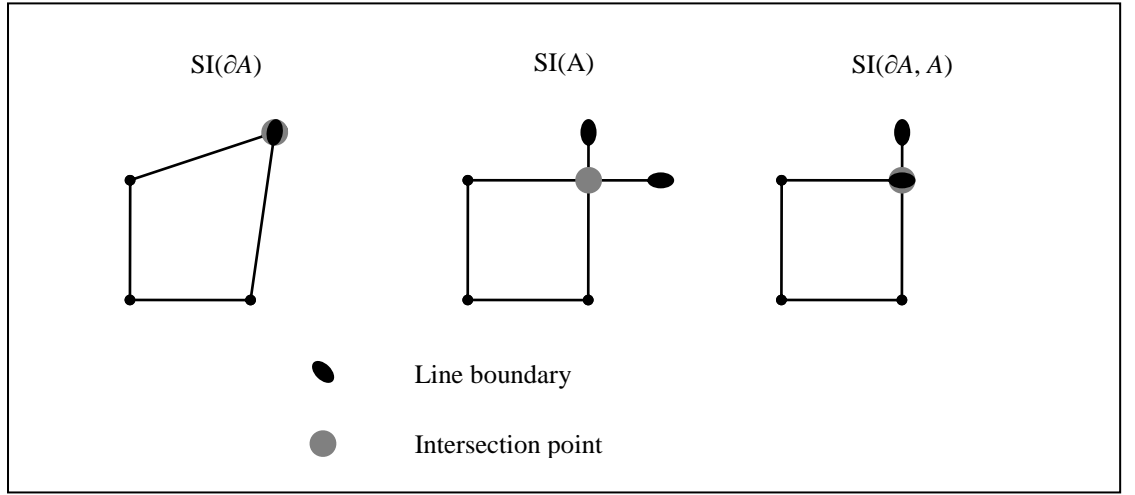

**Figure1**

2.3 Network topology

The networks of linear features discussed in this paper are based on the arc-node network model, let us then describe it using the 9-Intersection model. For simplicity reasons we use the notation $\{\varnothing, \neg\varnothing\}$ instead of the original notation $\{\varnothing, \neg\varnothing\}$ of the 9-Intersection model. When intersection between objects cannot be defined the notation "-" is used.

Regardless if a network is based on the planar or the nor-planar network model, the network elements that are considered to be connected must have their ends intersecting with connection points and with each other, reference and analysis of these models can be found in (Fischer, 2003).

Line ($L$) to point ($P$) valid intersections.

$$I_9(P, L) = \begin{Bmatrix} - & - & - \\ 0 & 1 & 0 \\ 1 & 1 & 1 \end{Bmatrix}$$

$$I_9(L, P) = \begin{Bmatrix} - & 0 & 1 \\ - & 1 & 1 \\ - & 0 & 1 \end{Bmatrix}$$

When lines $L_1$, $L_2$ that belong to the network are connected (adjoined).

$$I_9(L_1, L_2) = \begin{Bmatrix} 0 & 0 & 1 \\ 0 & 1 & 1 \\ 1 & 1 & 1 \end{Bmatrix}$$

When lines $L_1$, $L_2$ are disjoined.

$$I_9(L_1, L_2) = \begin{Bmatrix} 0 & 0 & 1 \\ 0 & 0 & 1 \\ 1 & 1 & 1 \end{Bmatrix}$$

In most implementations of networks of linear features Self-Intersection for any linear feature *A* is not permitted.

$$SI_L(A) = \{0 \quad 0 \quad 0\}$$

## 2.4 Topological errors and detection algorithms

Topological errors occur when the network elements are not connected as specified. For networks of the arc-node model a common error is when linear features that should be connected are disjoined due to bad digitization. Since the nodes are attached to the endpoints of the linear features, a trivial algorithm that detects these errors calculates the Euclidean distance of each node with every other. Such an algorithm based on distance criteria detects the nodes that are too close so as to indicate bad digitization.

Self-intersection is also considered invalid in most networks. The most important problem is that in many cases when $SI(\partial A) = 1$ routing problems are caused in applications that use the network. An algorithm that detects self-intersection errors is presented later in §7.

## 3 Geometrical reduction

The notion of reduction plays a key role in the designing of the algorithms presented in this paper. Through the reduction we build more efficient algorithms, which is a crucial matter especially when handling large databases. We define and discuss the notion of geometrical reduction that is used by most algorithms presented here. Geometrical reduction is the reduction of the spatial objects and relationships of a problem to objects easier to handle and relationships easier to compute. On this process the initial problem is transformed to a more easily solved problem.

The definitions and proof of geometrical reduction we provide in this section are based on the notion of mapping reduction (also called many-one reduction). Analysis of mapping reduction can be found in M. Sipser's book about computational complexity (Sipser, 2006).

**Geometrical reduction of object classes**
A computable function *g* reduces geometrically *n*-dimensional object class *A* to *d*-dimensional object class *B*, $A \leq_G B$, if for every object $o \in A \Leftrightarrow$ object $g(o) \in B$.

**Geometrical reduction of spatial relationships**
Spatial relationship *S*(*A*, *C*) is geometrically reducible to spatial relationship *P*(*B*, *D*), written $S(A, C) \leq_G P(B, D)$, if $A \leq_G B$ and $C \leq_G D$ and if *S*(*A*, *C*) is satisfied when *P*(*B*, *D*) is satisfied and *P*(*B*, *D*) is satisfied when *S*(*A*, *C*) is satisfied.

**Geometrical reduction theorem**
The problem *R* of deciding if a spatial relationship *P*(*B*, *D*) is valid in dataset *E* is mapping reducible to problem *V* of deciding if a spatial relationship *S*(*A*, *C*) is valid in dataset *J*, if $P(B, D) \leq_G S(A, C)$ and if there is a computable function *h* that decides *V*.

*Proof* : In order to prove the theorem it is only necessary to apply a computable function *f* so that for every instance *w* of *R* there is an instance *f(w)* of *V*, so that $w \Leftrightarrow f(w)$, let us build one.

Given input *w* that consists of the object classes *B*, $D \in E$, the empty object classes *A*, $C \in J$, functions $g_1$, $g_2$ that geometrically reduce the objects of *B*, *D* as specified in $P(B, D) \leq_G S(A, C)$ and function *h* that decides *V*. We build function *f*

function *f(B, D)*
       for every $b \in B$ create object $g_1(b)$ store in *A*
       for every $d \in D$ create object $g_2(d)$ store in *C*

       for every pair (*x*, *y*) where $x \in A$ and $y \in C$
       {
              if $h(g_1(x), g_2(y))$ is true
              then $S(g_1(x), g_2(y))$ is valid
              else
              $S(g_1(x), g_2(y))$ is not valid
       }

For every input *w* of *f* there is an output *f(w)* that is valid if *w* is valid, this comes as a conclusion since *P(B, D)* ≤$_G$ *S(A, C)*, so *w* ⇒ *f(w)*. Every output *f(w)* of *f* is produced from *w* and since *P(B, D)* ≤$_G$ *S(A, C)* then *f(w)* is valid if *w* is also valid, so *f(w)* ⇒ *w*.

As shown in theorem 5.22 in (Sipser, 2006) if *P* is decidable then *V* is also decidable.

## 4 Definition of hierarchical network structure and error detection problem

Before discussing error detection methods we must model the connection specifications in networks of linear features. For this reason we define categorization and hierarchy on the features of a studied dataset, so as to define a structure for the relationships among the features. This definition and the application of rules is a systematic way to model the connection specifications. The network elements are classified in categories according to their state or functionality. The rules that we apply model the obligation or the prohibition of each network element to be connected with other network elements of categories hierarchically higher or lower. The directive for the categorization process is that the more categories a feature is desired to be connected to, the larger weight we assign to it. This results a network with hierarchically connected features.

The idea for this categorization and the connection modeling comes from the hierarchy of roads in road networks. The elements of a road network are categorized according to their functionality and use. The ideal structure of a road network is formed by high capacity roads that act as the back bone of the network and all the lower capacity roads are connected to them. We may say that the network elements of higher categories are required to be connected with each other forming closed networks and the elements of the lower categories should be connected with elements of the same or higher categories. There are also categories of roads that the connection of their elements is prohibited, for instance motorways in general must not end up in unpaved and narrow roads.

The connection modeling and error detection, exactly as defined in this paper, may also be applied in networks designed for hierarchical routing algorithms. In these cases connection errors denote validation of the hierarchical structure of the network, which may result faulty execution of the algorithm. We may refer to the white paper (ESRI, 2005b) that describes the implementation of hierarchical routing in ArcGIS and the problems that may occur due to errors in the development of the network.

Let us define a network, which forms a connected graph and all its features are categorized in a finite number of *k* categories based on their weights, $k \in N$. For simplicity reasons the weights can be consecutive integers starting from 1, the smaller a weight is the higher category it symbolizes. Also we define variable $a \in N$ where $1 \leq a \leq k$.

The network with hierarchically connected linear features is defined by three rules.
1. Every feature with weight smaller than *a* must be connected, in both ends, with features of the same or higher category.
2. Features belonging in the highest and in the lowest category must not be connected.
3. Every feature with weight higher or equal to *a* must be connected, in one or two ends, with features of the same or higher category.

The first rule is applied in features that must form closed networks with other features of the same or higher category. When a feature is connected in both ends with feature of the same or higher category is part of a circle, or closed network. As an example we may refer to motorways and freeways in road networks.

The second rule is applied in features that their connection is prohibited. If there is any prohibition of connecting two categories they must follow the second rule. The application of the prohibition in the features of the higher and lower categories is indicative but also realistic. There are many types of networks where features of higher categories must not be connected with features of lower categories. For example in telecommunication networks, high capacity channels that interconnect the network service providers and are not directly connected to home networks.

The third rule is applied in features that must be connected with features of the same or higher category; the formation of a closed network is not required. Examples of features that must follow the third rule are communication channels that end in terminals or dead-end streets in road networks.

Connection problems regarding the maximum or minimum number of features that are allowed to be connected in each connection point are related with the capacity and the flow capabilities of the network elements, which is outside the topic of this paper and they will not be discussed.

**Detection of connection errors in networks with hierarchically connected linear features.**

Given a network of linear features $N$ that forms a connected graph of $|E|$ linear features $e_1\ldots e_j$ of corresponding weights $w_1\ldots w_j$ and $|V|$ connection points $v_1\ldots v_s$. Also an integer variable $a$, where $1 \leq a \leq max(w_j)$.

In which connection points for every feature $e_g$ of weight $w_g < a$ that is connected to there is no other feature $e_u$ connected so that $w_u \geq w_g$?

In which connection points there are connected features $e_l$ and $e_x$ of weights $w_l$ and $w_x$ so that $w_l = min(w_j)$ and $w_x = max(w_j)$ ?

Which feature $e_t$ of weight $w_t \geq a$ is not connected with feature $e_p$ connected so that $w_p \geq w_t$ ?

**Reduction of special cases to the problem**

The definition of the problem describes a simple and general case of the problem. Let us describe the reduction of more special and complex cases to the general problem. Processes that reduce relative problems to the above must have the following functions:

1. In networks that form more than one connected parts each connected part is considered as a different network.
2. In networks where the categorization is defined by more than one attribute the categorization can be redefined by combining the attributes. For example if we need to combine one attribute of $k$ categories and one of $g$ categories the result would be a new categorization of $k \times g$ categories.
3. In cases where the categories that must follow the first or the third rule have not consecutive weights, the variable $a$ is replaced by a set of integer numbers. In case that the weight of a feature belongs to the set the feature follows the third rule if not it follows the first.
4. In GIS the linear features are directed arcs. We can use this property to symbolize networks with directed features e.g. one way streets in road networks. If the direction of the network flow is the same as the direction of the arc the feature weight is 1, if it is the opposite direction the weight is -1 and in cases where the element is valid in both directions the weight is 0.
5. In directed networks where there are different specifications at the start and the end of the linear features the algorithms that solve the problem must consider the beginning and the end of each arc as two different features. Since each end of the line can be connected with different feature classes the algorithms must examine them separately. An example for this kind of specification is when there is the need to ensure that cities have access to motorways through a specific category of one way roads, so the start of these roads must be connected with city streets network and the end of the road with roads categorized as motorways.

As shown later in §7 the reductions of the problem can also be used for the detection of network flow problems.

## 5 Exhaustive search algorithm

A common structure for the attribute table of linear features is a two-dimensional table with one record per feature and at least three fields, the first field for the unique identification number of the feature, the second field for the identification number of the node that the digitized arc starts and the third field for the node that the arc ends. If the feature has weight this is stored in the fourth field of the table and other possible information in extra fields. Analysis of this structure may be found in (Fischer, 2003).

We present an algorithm that detects connection errors in datasets with attribute tables of this common structure. The algorithm performs exhaustive search in the table and outputs the features that participate in erroneous connections errors in a new dataset. The reason we discuss this algorithm is that it uses an extensively used dataset structure, so the method the algorithm uses may be applied in many modern projects that use datasets of this structure. The exhaustive search has bad computational performance $O(n^2)$, since every linear feature is compared with every other. The performance could be improved by indexing the attribute table, the analysis of such a method is out of the scope of this paper since we use a different approach for building efficient algorithm that detect connection errors.

Considering the structure of the attributes table and the network topology we may say that two linear features $L_i$ and $L_j$ of dataset SET0 are considered connected if and only if (SET0[$i$][2] or SET0[$i$][3]) == (SET0[$j$][2] or SET0[$j$][3]), where $i, j \in N$.

**Exhaustive search algorithm**

```
input:    SET0;                        // network attributes table
          SET_ERR;                     // empty dataset, for error storage
          int a;                       // variable a as defined on the problem
          int n = number of rows of SET0;
          int max = maximum category weight;
          int min = minimum category weight;

procedure EXHAUSTIVE_SEARCH
          for i = 1 to n               // the procedure is executed for every feature
          {
                    if SET0[i][4] == max or SET[i][4] == min
                    run procedure RULE2         // error detection of rule 2
                    then

                    if SET0[i][4] < a
                    run procedure RULE1         // error detection of rule 1

                    else
                    run procedure RULE3         // error detection of rule 3
          }
          halt

procedure RULE1
input: feature SET0[i]
          for j = 1 to n
          {
                    if i <> j
                    then                        // check of the start of the arc
                    if SET0[i][2] == (SET0[j][2] or SET0[j][3])
                    and if SET0[j][4] <= SET0[i][4]
                    set start = true
                                                // check of the end of the arc
                    if SET0[i][3] == (SETo[j][2] or SET0[j][3])
                    and if SET0[j][4] <= SET0[i][4]
                    set end = true
          }
          if start ∧ end = false          // check of connections in both ends
          SET0[i] to SET_ERR
          return

procedure RULE2
input: feature SET0[i]
          for j = 1 to n
          {                             // check if there is invalid connection
                    if discrete features SET0[i] and SET0[j] are connected
                    and if SET0[j][4] == max or SET[j][4] == min
                              and SET0[i][4] <> SET0[j][4]
                    store SET0[i] to SET_ERR
                    return
          }
          return
```

```
procedure RULE3
input: feature SET0[i]
      for j = 1 to n
           {                                    // check if there is valid connection
                if discrete features SET0[i] and SET0[j] are connected
                then
                if SET0[j][4] <= SET0[i][4]
                return
           }
           store SET0[i] to SET_ERR         // if no valid connection found store as error
           return
```

## 6 Spatial join algorithm

In this and the next sections we define algorithms that detect connection errors more efficiently than the exhaustive search. The algorithms use the notion of geometrical reduction in order to reduce the computation of the problem to the computation of spatial join in point feature datasets.

The spatial relationship of two connected linear features ($L_1$, $L_2$) can be geometrically reduced to the spatial relationship of their endpoints ($P_1$, $P_2$) at the connection point (node) of the network. The geometrical reduction we describe is the operation of creating endpoints from lines which is an easy operation for most GIS and it takes $O(n)$ time. The spatial relationship of two intersecting points is more easily computable and the algorithm is more efficient. For the detection of the spatial intersection we use spatial join and the dataset uses R-Tree indexing. As it will be analyzed in §8 R-Tree indexing improves the performance of the algorithms.

The geometrical reduction is $I(L_1, L_2) \leq_G I(P_1, P_2)$ for

$$I_9(L_1, L_2) = \begin{Bmatrix} 0 & 0 & 1 \\ 0 & 1 & 1 \\ 1 & 1 & 1 \end{Bmatrix}$$

$$I_9(P_1, P_2) = \begin{Bmatrix} - & - & - \\ - & 1 & 0 \\ - & 0 & 1 \end{Bmatrix}$$

A reduction instance is illustrated in figure 2.

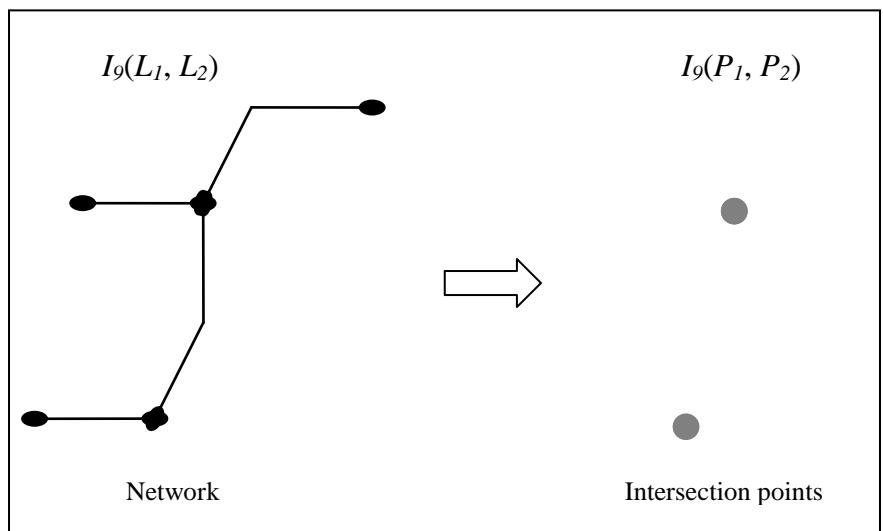

$I_9(L_1, L_2)$              $I_9(P_1, P_2)$

Network              Intersection points

**Figure 2**

For every pair of linear features that are connected there is a pair of endpoints that intersect and for every pair of intersecting endpoint there is a pair of connected linear features. In order to make the algorithms even more efficient the points that are created are stored in different datasets for each category of linear features and each dataset is being examined for legal or illegal connections only with the categories that is necessary not with the whole network. The start point and the end point of the features that follow the first rule will be stored in deferent datasets.

One more advantage of the algorithms is that it can be applied in features that belong to more than one datasets without the need for definition of linear topology.

We next present an algorithm for the general case of the problem. The initial dataset of linear features containing the network is the SET0; it may have the same structure as in the exhaustive search algorithm. The algorithm creates empty datasets of point features with one field in the attribute table named "CHECK". We name SETx the datasets for the endpoints of features with weight less than the variable $a$, as it was defined on the general case of the problem. We create two different datasets for the endpoints of the features with weight larger or equal to $a$, SETx_1 for the start point and SETx_2 for the endpoint. Variable x equals to the weight of the features of each category and $1 \leq x \leq 2k - a + 1$.

**Spatial join algorithm**

```
input:  SET0                              // Initial dataset
        SET1, SET2, ..... , SETa-1        // a - 1 empty datasets
        SETa_1, SETa_2, ..... , SETk_1, SETk_2   // 2(k - a +1) empty datasets
        SET_R2                            // rule 2 errors storage dataset
        SET_R13                           // rules 1,3 error storage dataset
        int a;                            // variable a as defined on the problem
        int k = number of categories;
        int n = number of linear features;

procedure EXHAUSTIVE_SEARCH
        run procedure REDUCTION
        then

        for i = 1 to k                    // the procedure is executed for every point dataset
        {
                if i < a
                run procedure RULE1 for SETi

                else
                run procedure RULE3 for SETi_1 and SETi_2
                then if i == k
                run procedure RULE2
        }
        halt

procedure REDUCTION
        for i = 1 to n                    // the procedure is executed for every feature
        {
                f = SET0[i][4]            //variable f defines the dataset we store endpoints
                if f < a
                Create StartPoint of SET0[i] and store in SETf_1
                Create EndPoint of SET0[i] and store in SETf_2
                else
                Create StartPoint of SET0[i] and store in SETf
                Create EndPoint of SET0[i] and store in SETf
                return
        }
```

```
procedure RULE1(i)
        for j = 1 to i
        {
                "UPDATE SETi SET SETi.CHECK = 1
                WHERE SpatialJoin (SETi, SETj) is satisfied"
        }

        store in SET_R13 the features
        "SELECT * FROM SETi WHERE SETi.CHECK <> 1"
        return

procedure RULE2
                store in SET_R2 the features
                "SELECT * FROM SET1 WHERE SpatialJoin (SET1, SETk_1) is satisfied"
                "SELECT * FROM SET1 WHERE SpatialJoin (SET1, SETk_2) is satisfied"
                "SELECT * FROM SETk_1 WHERE SpatialJoin (SETk_1, SET1) is satisfied"
                "SELECT * FROM SETk_2 WHERE SpatialJoin (SETk_2, SET2) is satisfied"
                return

procedure RULE3(i)
        for j = 1 to i
        {
                "UPDATE SETi_1 SET SETi_1.CHECK = 1
                WHERE SpatialJoin (SETi_1, SETj) is satisfied"

                "UPDATE SETi_2 SET SETi_2.CHECK = 1
                WHERE SpatialJoin (SETi_2, SETj) is satisfied"

        }

        store in SET_R13 the features
        "SELECT * FROM SETi_1 WHERE SETi_1.CHECK <> 1"
        "SELECT * FROM SETi_2 WHERE SETi_2.CHECK <> 1"
        return
```

## 7 Line Self-Intersection and flow error detection

Based on the geometrical reduction that was used in the algorithm of the previous section and the special cases of the problem that was discussed in §4, we present two algorithm that detect network flow errors and line Self-Intersection. For simplicity reasons, we consider that the flow in each feature is the same as the direction of the arc that denotes it.

The Point no flow algorithm detects the connection points in directed networks in which network flow is blocked. These errors occur when there are nodes that the traffic is directed to them but not directed from them or when there are inaccessible nodes that the traffic is only directed from them. The endpoints that are created from the geometrical reduction of the arcs of the network are categorized in two categories and stored in datasets SET1 and SET2. The criterion for the categorization is whether the traffic on the arc starts from or ends to the endpoint. Every point of the first category must intersect with at least one point of the second category, so that the traffic flow may be lead to and out of every node. An instance of the problem and the output of the algorithm are illustrated in figure 3.

**Point no flow algorithm**

```
input :   SET0                                    // Initial dataset
          SET1, SET2                              // 2 empty datasets
          SET_ERR                                 // empty dataset, for error storage
          int n = number of linear features;

procedure POINT_NO_FLOW
          run procedure REDUCTION

          store in SET_ERR the features
          "SELECT * FROM SET1 WHERE SpatialJoin (SET1, SET2) is not satisfied"

          store in SET_ERR the features
          "SELECT * FROM SET2 WHERE SpatialJoin (SET2, SET1) is not satisfied"
          halt

procedure REDUCTION
          for i = 1 to n                          // the procedure is executed for every feature
          {
                    Create StartPoint of SET0[i] and store in SET1[i]
                    Create EndPoint of SET0[i] and store in SET2[i]
                    return
          }
```

**Figure 3**

The Line Self-Intersection algorithm detects cases where SI(∂A) = 1. In these cases both the endpoints of a linear feature intersect. The endpoints categorization is similar to the previous algorithm.

**Line Self-Intersection algorithm**

```
input :   SET0                             // Initial dataset
          SET1, SET2                       // 2 empty datasets
          SET_ERR                          // empty dataset, for error storage
          int n = number of linear features;

procedure POINT_NO_FLOW
          run procedure REDUCTION

          for j = 1 to n
          {
                  if SpatialJoin (SET1[j], SET2[j]) is satisfied
                  store SET1[j] in SET_ERR
          }
          halt

procedure REDUCTION
          for i = 1 to n                   // the procedure is executed for every feature
          {
                  Create StartPoint of SET0[i] and store in SET1[i]
                  Create EndPoint of SET0[i] and store in SET2[i]
                  return
          }
```

## 8 Complexity analysis

All the algorithms use as input the linear features of the network, let $n$ be the cardinality of the edges of the studied network. As shown the exhaustive algorithm needs $O(n^2)$ time for search.

The geometrical reduction of the linear features to endpoints has linear complexity $\Theta(n)$, as it is executed once for each of the linear features. The algorithms that use geometrical reduction execute spatial join over the reduced datasets in order to detect the spatial relationships of the features. Modern GIS use methods of improving the performance of data search by building spatial indexes that makes spatial joins more efficient, a presentation can be found in (Jacox and Samet, 2005). One widely used method is the use of R-Trees that improves considerably spatial join operations as shown in (De Berg et al, 2003) and in many cases it can even achieve logarithmic search complexity as described in (Göbel, 2007). In the worst case scenario of spatial join between two point features the operation takes linear time. This can happen when the indexed datasets have no intersecting objects. So the spatial object processes in the algorithms of this paper have linear or better time complexity.

In the point no flow algorithm the two spatial joins accept as input $n$ point features, so the complexity of the two processes is at worst linear. The Line Self-Intersection algorithm executes $n$ times spatial joins for two point features each time, the complexity of this process is $\Theta(n)$. Since the geometrical reduction of the linear features has greater or equal complexity to the spatial join processes, we may say that the overall performance of these two algorithms is mainly affected by geometrical reduction. The complexity is $O(n)$ for the point no flow and $\Theta(n)$ for the Line Self-Intersection.

For the estimation of the complexity of the algorithm based on spatial join that solves the general case of the problem, we have to estimate the complexity of the spatial joins that are executed each time. The average number of features per category is $n_{av} = n / k$, without loss of generality we use this distribution of features per category for the estimation of the complexity of the algorithm. Since the dataset of each category executes spatial joins with itself and all the datasets of higher categories, the total of the executed spatial join processes is $k^2 / 2$ if we apply the first rule to all categories and $k^2$ if we apply the third rule. Based on these the worst case scenario is when we use $n$ categories, one for each feature and we apply the third rule resulting $O(n^2)$ complexity. But this scenario implements a totally unrealistic network with no practical use and no representation of any real world network. In reality, network features are categorized in a small number of $k$ categories where $k << n$. For instance, in road networks are usually used ten categories at most. So in realistic implementations of networks the worst case scenario complexity is $k^2 \times n_{av} = k \times n = O(n)$. Since all the rest processes of the algorithm have linear complexity, the complexity of the whole algorithm is linear.

## 9 Why we cannot build a general method for algorithmic error correction

Since we may build procedures that detect errors, can we build procedures that algorithmically correct them?

There have been many attempts from researchers to discover a theoretical framework that would result an affirmative answer to this question. In this section we prove the contrary.

The algorithmic correction of maps is a special case of a more general problem, the algorithmic correction of software bugs and an empirical answer to this question would be "in general, no". It is very difficult to build an algorithmic procedure that would take into account all the necessary parameters and lead to the desired result by correcting all the instances of errors in a piece of software or in a map. As we will show next such a problem is undecidable and it may be solved only in simple instances. On the proof, we use the Turing Machine (TM) as a model of general purpose computer and define the problem of algorithmic correction of software errors as an equivalent computational problem of algorithmic correction of the input string of a Turing Machine.

Let $M$ be a Turing Machine that accepts input $w$. If $w$ is build under certain specifications $M$ prints a message of acceptance and then halts. In case $w$ does not comply with specifications, $M$ might not halt or have unpredicted behavior. Now let $N$ be another Turing Machine that examines any given input $w$. If $w$ is incorrect according to the specifications and the description of $M$, then $N$ corrects $w$ so $M$ can run $w$ properly and then halt.

Formally:

INPUT-CORRECTION$_{TM}$ = {<$M$, $N$, $w$> | $M$ and $N$ are TM, $N$ modifies $w$ according to the description of $M$ so that $M$ halts on input of modified $w$}

In order to prove the undecidability of this problem it is necessary to define another undecidable problem the HALT problem. Undecidability implies that it cannot be decided by a Turing Machine and an algorithm that solves it for any given input $w$ cannot be built.

HALT$_{TM}$ = {< $M$, $w$> | $M$ is a TM and $M$ halts on input $w$}

**Theorem**. The INPUT-CORRECTION problem is undecidable.

*Proof*: Turing Machine $N$ modifies any input $w$, according to the description of $M$, so $M$ can run $w$ and halt. If $w$ is correct $N$ prints "YES", $M$ runs $w$ produces an output and halts. If $w$ is incorrect and $M$ cannot halt while running $w$, $N$ corrects $w$ and prints "YES" then $M$ runs $w$ produces an output and halts.

Let $H$ be a machine that simulates the operation of both $M$ and $N$, then $H$ halts on any given input and the HALT problem is decidable for $H$ and any given input $w$. This is a contradiction since the HALT problem is undecidable.

### 9.1 A more intuitive approach

Let's look into more practical perspective of this issue. The specifications of every spatial network define the valid spatial relationships of the network elements as well as the valid attributes of every element. Attributes and spatial relationships cannot be examined independently since the validation of both of them ensures network consistency. Let $A$ be the set of all the possible valid combinations among the valid spatial relationships and the valid attributes of the network elements. Let $B$ be the set of all the possible combinations of spatial relationships and attributes that are considered invalid according to specifications.

Let $f$ be an algorithmic function that may correct the errors in a given network. Then $f$ accepts as input one element of $B$ and outputs one element of $A$. This means that each element of B must correspond exactly to one element of $A$; $f$ must be 1-1 function or many-to-one (n-1). But this can happen only to very special and simple cases; it cannot be applied in general. If an element of *B* corresponds to many elements of *A* then *f* may not be defined.

On the same way we may define sets *A* and *B* for any other software structure. An 1-1 or *n*-1 function among software with bugs and error free software may be found only in very simple instances.

**An example that concerns spatial networks.**

Let *N* be a non-planar network of linear features. Let *a* and *b* be two elements of *N* where the edge of *a* crosses the middle of the edge of *b*. Suppose that the person who draw *a* and *b* neglected to apply on them the necessary parameters that denote their structure and state. We may consider the following possible cases.

- The edge of *a* is on ground level and the edge of *b* overpasses *a*, edge *b* represents a bridge.
- The edge of *a* is on ground level and the edge of *b* underpasses *a*, edge *b* represents a tunnel.
- Both edges are not on ground level and edge *b* overpasses edge *a*, multilevel traffic junction.
- Both edges are in ground level, *a* and *b* represent a crossroad. Both edges must split on the intersection point and form a proper crossroad.

There is no way how we can build an algorithm that will decide which of the four cases corresponds to the given instance, so as to correct the error.

If we consider a special case of network instance in which we may find an 1-1 function between sets *A* and *B*, we could perform algorithmic correction of a specific error case. As it is easy to see this error case will be very simple. For instance, if we consider a network where each element that participates in a specific computable spatial relationship must be characterized by an attribute of a single value. Then an algorithmic process that assigns this value to the attributes of the appropriate elements may be build. Nevertheless this would be of minor importance in the complex task of error correction.

## 10 Conclusions

As we have shown, geometrical reduction and R-Tree indexing decrease the complexity of spatial algorithms. Therefore the algorithms we present can be used for the efficient detection of connection errors in networks of linear features. The elements of network we give as input to the algorithms must be hierarchically categorized. Through the hierarchical categorization we model the connection specifications of the network elements. The connection specifications regard the obligation or prohibition of connection among features of certain functionality, as well as the prohibition of line self-intersection and the avoidance of flow problems.

At last, the important corrections of detected errors in maps and in general in computer software have to be manual and of course by experienced personnel.